\documentclass{aa}
\usepackage{graphicx}
\usepackage{amssymb}
\usepackage{amsmath}
\usepackage{url}
\usepackage{natbib}
\usepackage{amssymb}
\usepackage{amsmath}
\begin{document}
   \title{CARS: The CFHTLS-Archive-Research Survey \\ III. First
     detection of cosmic magnification in samples of normal high-$z$
     galaxies\thanks{Based on observations obtained with
       MegaPrime/MegaCam, a joint project of CFHT and CEA/DAPNIA, at
       the Canada-France-Hawaii Telescope (CFHT) which is operated by
       the National Research Council (NRC) of Canada, the Institut
       National des Sciences de l'Univers of the Centre National de la
       Recherche Scientifique (CNRS) of France, and the University of
       Hawaii. This work is based in part on data products produced at
       TERAPIX and the Canadian Astronomy Data Centre as part of the
       Canada-France-Hawaii Telescope Legacy Survey, a collaborative
       project of NRC and CNRS.}} \titlerunning{Cosmic magnification
     with LBGs}

   \subtitle{}

   \author{H. Hildebrandt\inst{1}
     \and L. van Waerbeke\inst{2}
     \and T. Erben\inst{3}
   }

   \offprints{H. Hildebrandt}

   \institute{Leiden Observatory, Leiden University, Niels Bohrweg 2,
     2333CA Leiden, The
     Netherlands\\ \email{hendrik@strw.leidenuniv.nl} \and
University of British Columbia, Department
     of Physics and Astronomy, 6224 Agricultural Road, Vancouver,
     B.C. V6T 1Z1, Canada\\ \and      Argelander-Institut f\"ur Astronomie, Auf dem H\"ugel 71, 53121
     Bonn, Germany
}

   \date{Received ; accepted }

  \abstract {Weak gravitational lensing (WL) has been established as
    one of the most promising probes of cosmology. So far, most
    studies have exploited the shear effect of WL leading to coherent
    distortions of galaxy shapes. WL also introduces coherent
    magnifications.}{We want to detect this cosmic magnification
    effect (coherent magnification by the large-scale structure of the
    Universe) in large samples of high-redshift galaxies selected from
    the Deep part of the Canada-France-Hawaii-Telescope Legacy Survey
    (CFHTLS).}{Lyman-break galaxies (LBGs) selected by their colours
    to be at $z$=2.5-5, are used as a background sample and are
    cross-correlated to foreground lens galaxies, which are selected
    by accurate photometric redshifts (photo-$z$'s). The signals of
    LBGs in different magnitude bins are compared to predictions from
    WL theory. An optimally weighted correlation function is estimated
    by taking into account the slope of external LBG luminosity
    functions.}{For the first time, we detect cosmic magnification in
    a sample of normal galaxies. These background sources are also the
    ones with the highest redshifts so far used for WL
    measurements. The amplitude and angular dependence of the
    cross-correlation functions agree well with theoretical
    expectations and the lensing signal is detected with high
    significance. Avoiding low-redshift ranges in the foreground
    samples which might contaminate the LBG samples we can make a
    measurement that is virtually free of systematics. In particular,
    we detect an anti-correlation between faint LBGs and foreground
    galaxies which cannot be caused by redshift
    overlap.}{Cross-correlating LBGs (and in future also photo-$z$
    selected galaxies) as background sources to well understood
    foreground samples based on accurate photo-$z$'s will become a
    powerful cosmological probe in future large imaging surveys.}


   \maketitle
%
%________________________________________________________________

\section{Introduction}
Many studies have shown so far that images of faint background
galaxies are coherently distorted by the gravitational lensing effect
of the large-scale structure in the line of sight. This cosmic shear
effect \citep[for recent reviews
  see]{2008PhR...462...67M,2008ARNPS..58...99H} has been identified to
be one of the most promising approaches to study the properties of
dark matter and dark energy
\citep{2006astro.ph..9591A,2006ewg3.rept.....P}. It relies on the
assumption that the ellipticities of background galaxies are
intrinsically randomly distributed. Cosmic shear introduces tiny,
coherent distortions to this random distribution, which can be
measured \citep[see e.g. ][]{2007MNRAS.381..702B,2008A&A...479....9F}.

Besides this shear effect of gravitational lensing the images of
background galaxies are also subject to magnification by the
large-scale structure of the Universe. However, no simple assumption
about the intrinsic distribution of the fluxes of the background
galaxies can be made. Rather, this distribution has to be measured
from the data by averaging over large areas of the sky.

The magnification effect of gravitational lensing to first order
depends on the convergence only, which is the projected mass along the
line-of-sight. This effect is geometrical in nature enlarging the
solid angle behind masses. This leads to two distinct effects, a
dilution of the source density and a magnification of their fluxes
since lensing conserves the surface brightness. Now it depends on the
intrinsic distribution of the fluxes whether the angular sky positions
of the background galaxies of a given observed flux are positively or
negatively cross-correlated to the angular positions of the foreground
masses. If there are many more faint background galaxies than bright
ones, i.e. steep magnitude number counts, the magnification wins over
the dilution and a positive angular cross-correlation is expected. In
the case of shallow number counts the dilution becomes dominant and a
negative angular cross-correlation is expected. This general effect of
gravitational lensing is called magnification bias and in the case of
the large-scale structure of the Universe as the lens it is referred
to as cosmic magnification.

Measurements of the cosmic magnification signal can be interpreted in
a similar way as signals of galaxy-galaxy lensing \citep[see
  e.g. ][]{1996ApJ...466..623B,2004ApJ...606...67H,2006MNRAS.368..715M,2007ApJ...669...21P}. Thus,
cosmic magnification can be directly employed to study the dark matter
environment of galaxies constraining the galaxy bias, the total matter
density, and the normalisation of the dark matter power spectrum.

The biggest problem in using magnification as a cosmological probe is
to cleanly separate the fore- and the background samples in
redshift. If there is some redshift overlap between the samples -
i.e. if the lens and source galaxies are physically close - they see
the same dark matter field and will hence cluster with respect to each
other. Then the angular cross-correlation signal is not a pure lensing
signal anymore but it is contaminated by physical cross-correlations.

Due to this complication the effect has only been convincingly
measured with high significance in the Sloan Digital Sky Survey (SDSS)
by \cite{2005ApJ...633..589S} using optically selected quasars as
background sources \citep[see also ][ for a different estimator of the
  cosmic magnification signal with the same
  data]{2009arXiv0902.4240M}. For the long and controversial history
of such measurements of quasar-galaxy cross-correlation we refer the
reader to the references within that paper.

Here we present the first measurement of the same effect on normal
galaxies which have a much higher surface density on the sky than
quasars and can thus lead to much more accurate results. The
Lyman-break technique allows for the selection of clean samples of
high-redshift star-forming galaxies from optical data. In
\cite{2009A&A...498..725H} we presented the largest survey of these
galaxies to date. More than 80\,000 LBG candidates with redshifts
$z=2.5-5$ were selected from the data of the Deep part of the CFHTLS
and their clustering properties were measured. The same samples are
used here to detect cosmic magnification by cross-correlating them to
foreground galaxies.

In Sect.~\ref{sec:theory} we review the theoretical framework which is
necessary to interpret the measurements. The data analysis is covered
in Sect.~\ref{sec:data} and the results are presented in
Sect.~\ref{sec:results}. Conclusions and an outlook to future
applications of this method are given in
Sect.~\ref{sec:conclusions}. Throughout the paper we assume
$H_0=70\frac{\rm km}{\rm s\,Mpc}$, $\Omega_{\rm m}=0.3$,
$\Omega_\Lambda=0.7$, and $\sigma_8=0.8$, and we use AB magnitudes.

\section{Theoretical framework}
\label{sec:theory}
Let $N_0(>f)$ be the unlensed cumulative number counts of background
galaxies with fluxes larger than $f$. For simplicity we assume that
all background galaxies are at the same redshift. Foreground
structures will lead to a magnification $\mu$ at a particular position
on the sky. The lensed cumulative number counts are related to the
unlensed ones in the following way \citep{2001PhR...340..291B}:
\begin{equation}
  N(>f)=\mu^{-1}\,N_0(>\mu^{-1}f)\,,
\end{equation}
where the first $\mu$ corresponds to the dilution of the sample due to
the enlargement of the solid angle behind the lens and the second
$\mu$ corresponds to the brightening due to the enlargement of the
sources.

We assume that $N_0(f)$ can be approximated by a power-law:
\begin{equation}
\label{eq:flux_power_law}
  N_0(>f)=A\,f^{-\alpha}\,,
\end{equation}
with $A$ being the amplitude and $\alpha$ being the slope of the
power-law. The lensed cumulative number counts then become:
\begin{equation}
\label{eq:lensed_nc}
  N(>f)=\mu^{\alpha-1}N_0(>f)\,,
\end{equation}
Thus, it depends on the slope $\alpha$ of $N_0(>f)$ whether the
surface density of background sources is increased or decreased near
lenses where $\mu>1$. In the weak-lensing regime $\mu$ is close to
unity so that we can write $\mu=1+\delta\mu$ with $\delta\mu\ll1$ and
a Taylor expansion yields
\begin{equation}
\label{eq:mu_taylor}
  \mu^{\alpha-1}\approx1+(\alpha-1)\delta\mu\,.
\end{equation}

Using magnitudes instead of fluxes (i.e. substituting
$m=-2.5\log(f)+{\rm const.}$) it can easily be shown that $\alpha$ is
related to the differential magnitude numbercounts:
\begin{equation}
  2.5\frac{\mathrm{d}\log n(m)}{\mathrm{d}m}=\alpha\,,
\end{equation}
with $n(m)$ being the number counts of galaxies with magnitudes in the
interval $[m,m+\mathrm{d}m]$. By measuring the logarithmic slope of
the magnitude number counts we can predict over- or under-densities of
background galaxies induced by lensing.

Under the assumption of a linear biasing factor $b$ for the foreground
galaxies, the angular cross-correlation between these foreground
lenses and the background sources, $w_{\rm sl}(\theta)$, is related to
the angular cross-correlation between magnification and matter density
contrast, $w_{\mu\delta}(\theta)$ by:
\begin{equation}
\label{eq:sl_corr}
  w_{\rm sl}(\theta)=(\alpha-1)\,b\,w_{\mu\delta}(\theta)\,.
\end{equation}
For this result we employed
Eqs.~\ref{eq:lensed_nc}~\&~\ref{eq:mu_taylor}, the definition of the
cross-correlation function, and the assumption of a linear relation
between matter- and galaxy-density.

We calculate $w_{\mu\delta}(\theta)$ as described in
\cite{2001PhR...340..291B}:
\begin{multline}
\label{eq:mu_cross_delta}
  w_{\mu\delta}(\theta)=\frac{3H_0^2\,\Omega_{\rm m}}{c^2} \int_0^{\chi_{\rm
      H}}\mathrm{d}\chi'\,f_K(\chi)\,W_{\rm s}(\chi')\,G_{\rm
    l}(\chi')\,a^{-1}(\chi')\:\times
  \\ \int_0^\infty\frac{k\,\mathrm{d}k}{2\pi}\,P_\delta(k,\chi')\,J_0[f_K(\chi')\,k\,\theta]\,,
\end{multline}
with $\chi$ being the comoving distance, $f_K$ being the comoving
angular diameter distance, $W_{\rm s}(\chi)=\int_\chi^{\chi_{\rm
    H}}\mathrm{d}\chi' G_{\rm
  s}(\chi')\frac{f_K(\chi'-\chi)}{f_K(\chi')}$ being the weight
function of the sources, $G_{\rm l/s}$ being the normalised distance
distribution of the lenses/sources, $a$ being the scale factor,
$P_\delta$ being the matter power spectrum, and $J_0$ being the
0th-order Bessel function of the first kind.

In order to measure the signal of Eq.~\ref{eq:sl_corr} from data it is
of utmost importance to cleanly separate the sources from the
lenses. Otherwise physical cross-correlations will swamp the tiny
signal and make an interpretation in the framework of weak
gravitational lensing impossible because these physical
cross-correlation are typically larger than the cosmic magnification
signal by an order of magnitude.

\section{Data analysis}
\label{sec:data}
\subsection{The dataset}
The data used in this study are taken from the CFHTLS-Deep Survey, an
imaging survey with MEGACAM@CFHT in the filters $ugriz$ in four
independent fields of 1 square degree each. In the framework of the
CARS (CFHTLS-Archive-Research Survey) project we have collected all
publicly available data until July 21, 2008. The data reduction is
carried out with the THELI imaging reduction pipeline
\citep{2005AN....326..432E} and is described in detail in
\cite{2009A&A...493.1197E} and \cite{2009A&A...498..725H}.

Multi-colour catalogues are created with \emph{SExtractor}
\citep{1996A&AS..117..393B} in dual-image mode from images convolved
to the same seeing. Photo-$z$'s for all objects are estimated with a
modified version of the \emph{BPZ} code \citep{2000ApJ...536..571B}
including a correction for galactic extinction \citep[with the maps of
][]{1998ApJ...500..525S}, a re-calibration of the photometric
zeropoints and the template set with the help of spectroscopic
redshifts \citep[see ][]{2004_Capak_PhDT}, and a modified prior. For
details on the catalogue creation see \cite{2009A&A...498..725H}.

\subsection{The LBG catalogues}
In \cite{2009A&A...498..725H} we describe how we select large samples
of LBGs from these data. Simulations are set up to identify regions in
two-colour-space where high-redshift sources can be selected with high
efficiency and low contamination. In this way we select $\sim34\,000$
$u$-dropouts at $z\sim3.2$, $\sim36\,000$ $g$-dropouts at $z\sim3.8$,
and $\sim10\,000$ $r$-dropouts at $z\sim4.7$. The faintest LBGs that
can be selected in that way from these data have measured total
magnitudes of $r=27.6$, $i=27.8$, and $z=27.8$ for the $u$-, $g$-, and
$r$-dropouts, respectively.

\begin{figure}
\centering
\resizebox{\hsize}{!}{\includegraphics{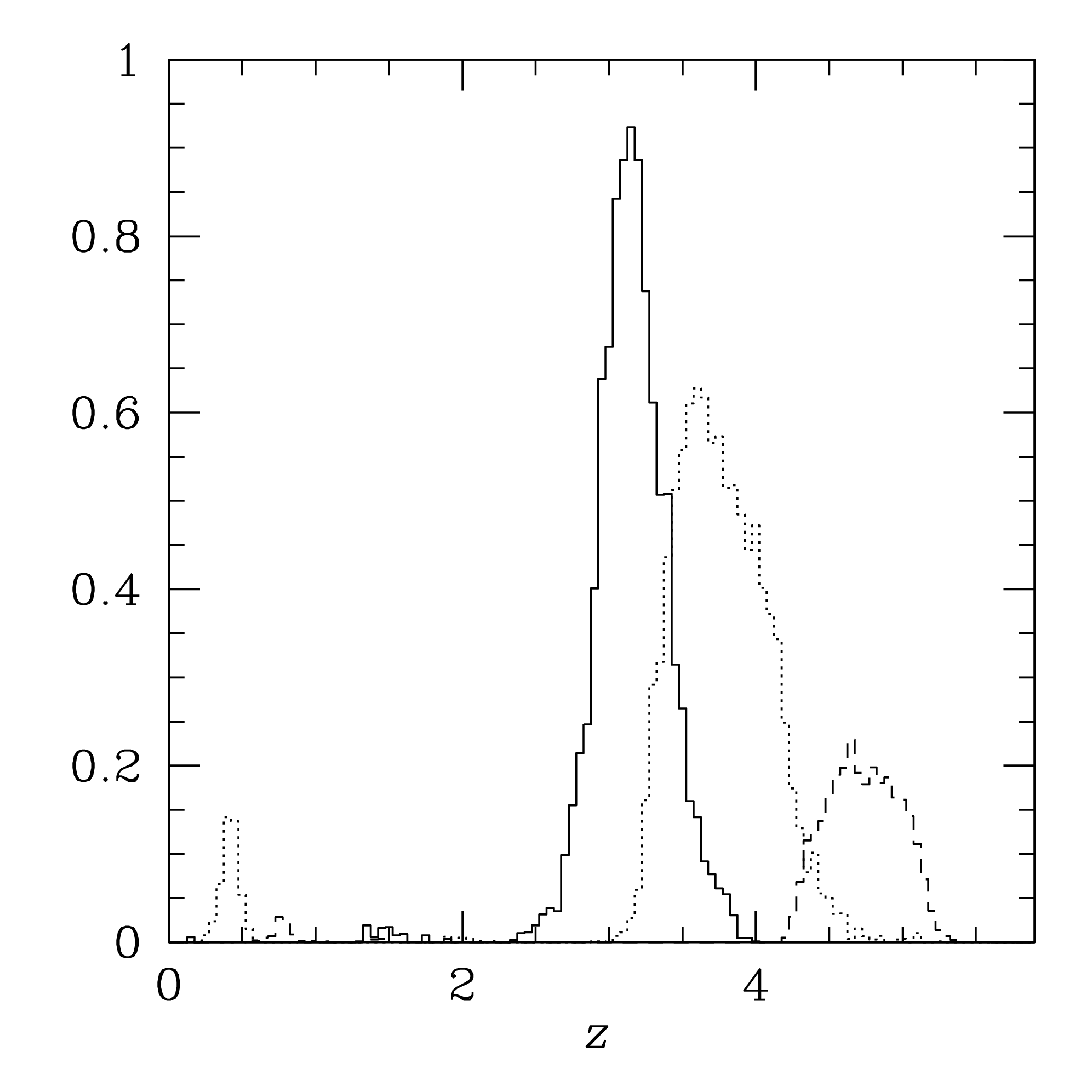}}
\caption{Simulated redshift distributions of the three dropout
samples
  ($u$-dropouts: \emph{solid}, $g$-dropouts: \emph{dotted},
  $r$-dropouts: \emph{dashed}; arbitrarily normalised but with correct
  relative fractions). The simulations are based on templates from the
  library of \cite{1993ApJ...405..538B}. See \cite{2009A&A...498..725H} for a
  detailed description of the simulations.}
\label{fig:z_dist}
\end{figure}

The simulated redshift distributions are displayed in
Fig.~\ref{fig:z_dist}. These suggest that the $u$-dropout sample is
essentially free of any low-$z$ contamination \footnote{There might be
  some contamination from galaxies at $z\sim1.5$ but the fraction
  should be very small with $\sim1\%$ according to our simulations.},
whereas the $g$-dropout sample is contaminated by a small fraction
($\sim4\%$) of low-$z$ galaxies with redshifts $0<z<0.5$, and the
$r$-dropout sample is contaminated very slightly ($\sim2.5\%$) by
galaxies with redshifts $0. 5<z<1.0$. One of the big advantages of
using LBGs as background sources is that we know the redshifts of
the possible contaminants. In the remainder of the paper we try to
avoid these redshift regions, which might be affected by some small
amount of contamination, in our foreground lens samples in order not
to mix the lensing signal with a signal from physical
cross-correlations.

For a given LBG background sample we approximate $G_{\rm s}$ of
Eq.~\ref{eq:mu_cross_delta} by a Dirac delta-function at the mean
redshift of the LBGs. This approximation is valid because the comoving
distance - the quantity on which the lensing signal depends - does not
change appreciably over the range where the LBG redshift distribution
is different from zero, i.e. the distributions in comoving distance
are very narrow for the LBGs (in contrast to the redshift
distributions).

\begin{figure*}
\centering
\includegraphics[width=\hsize]{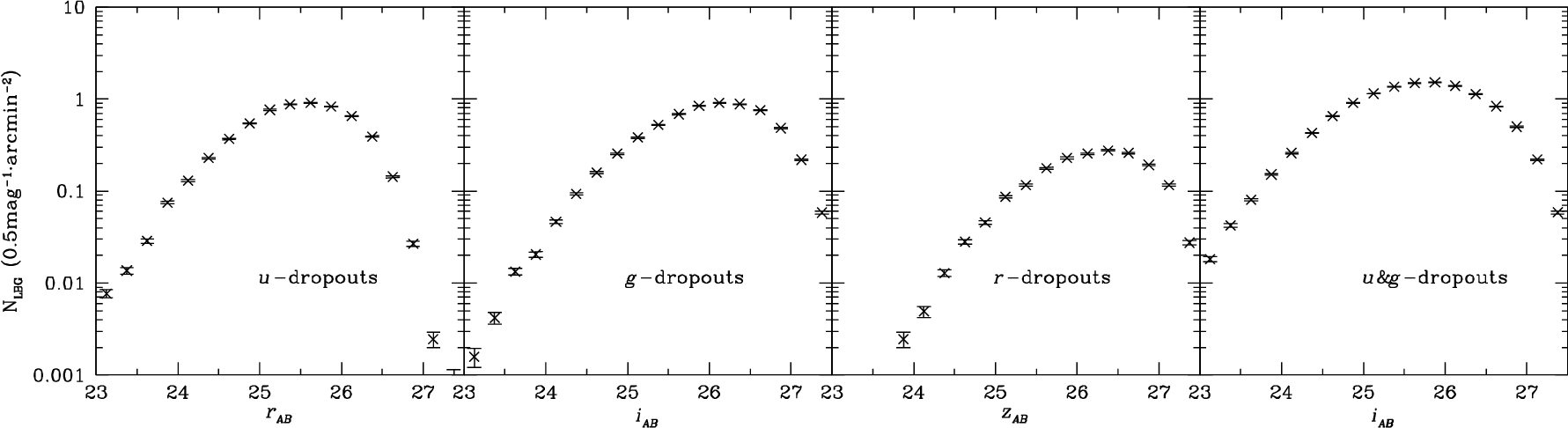}
\caption{Number counts of the three dropout samples and of the
  combined $u$\&$g$-dropout sample.}
\label{fig:numbercounts}
\end{figure*}

As described in Sect.~\ref{sec:theory}, the amplitude of the cosmic
magnification signal in the angular cross-correlation function of low-
and high-$z$ galaxies depends on the slope of the number counts of the
background sample. The number counts of the three samples and of the
combined $u$\&$g$-dropout sample are shown in
Fig.~\ref{fig:numbercounts}. For fainter magnitudes incompleteness
sets in. While this incompleteness does not bias the measurement of
the cross-correlation function (to first order), it prevents a
measurement of the slope of the number counts at the faint end, which
is necessary to carry out the theoretical predictions.

However, the number counts are closely related to the luminosity
function (LF). For a complete sample of galaxies in a thin redshift
slice the two curves are related to each other in magnitude by the
distance modulus and in amplitude by the volume normalisation. Thus,
using an external LBG-LF that has been properly corrected for
incompleteness one can predict the slope of the number counts. The
volume normalisation does not play a role here since we are only
interested in the slope.

The LBG-LF, $\Phi(M)$, was precisely measured in several studies at
$z\sim3$ \citep{1999ApJ...519....1S,2006ApJ...642..653S}, $z\sim4$
\citep{1999ApJ...519....1S,2004ApJ...611..660O,2004ApJ...600L.103G,2006ApJ...642..653S,2006ApJ...653..988Y,2007ApJ...670..928B},
and $z\sim5$
\citep{2003PASJ...55..415I,2007MNRAS.376.1557I,2004ApJ...611..660O,2004ApJ...600L.103G,2006ApJ...653..988Y}.\footnote{
  There is, however, some discrepancy between these literature
  measurements. A proper measurement of the LBG-LF from the
  CFHTLS-Deep data themselves is under way (van der Burg et al. in
  preparation). Certainly it is better to calibrate the data
  internally and such a LF estimate would make the external
  calibrators become redundant.}  Assuming a Dirac delta-function for
the redshift distribution of the LBGs, the slope of the number counts
of a complete sample of LBGs equals the slope of the luminosity
function. The parameter $\alpha$ introduced in Sect.~\ref{sec:theory}
can thus be expressed as
\begin{equation}
\label{eq:alpha}
\alpha(m)=2.5\,\mathrm{d}\log\mathrm{n}(m)/\mathrm{d}m=2.5\,\mathrm{d}\log\Phi(M)/\mathrm{d}M\,,
\end{equation}
with $m=M+DM+K$ and $DM$ being the distance modulus and $K$ being the
$K$-correction.

For the theoretical predictions we choose different LFs by
\cite{1999ApJ...519....1S}, \cite{2006ApJ...642..653S}, and
\cite{2007ApJ...670..928B} spanning a range of faint-end slopes and
parametrised in the way described by \cite{1976ApJ...203..297S}:
\begin{multline}
\Phi(M)\mathrm{d}M=0.4\,\Phi^*\ln(10)\left[10^{0.4(M_*-M)}\right]^{\alpha_{\rm LF}+1}\\
\times\exp[-10^{0.4(M_*-M)}]\mathrm{d}M\,.
\end{multline}
Note that $\alpha(m)$ of Eq.~\ref{eq:alpha} approaches the value of
the Schechter LF parameter $\alpha_{\rm LF}$ for faint magnitudes.

\begin{figure*}
\centering
\includegraphics[width=\hsize]{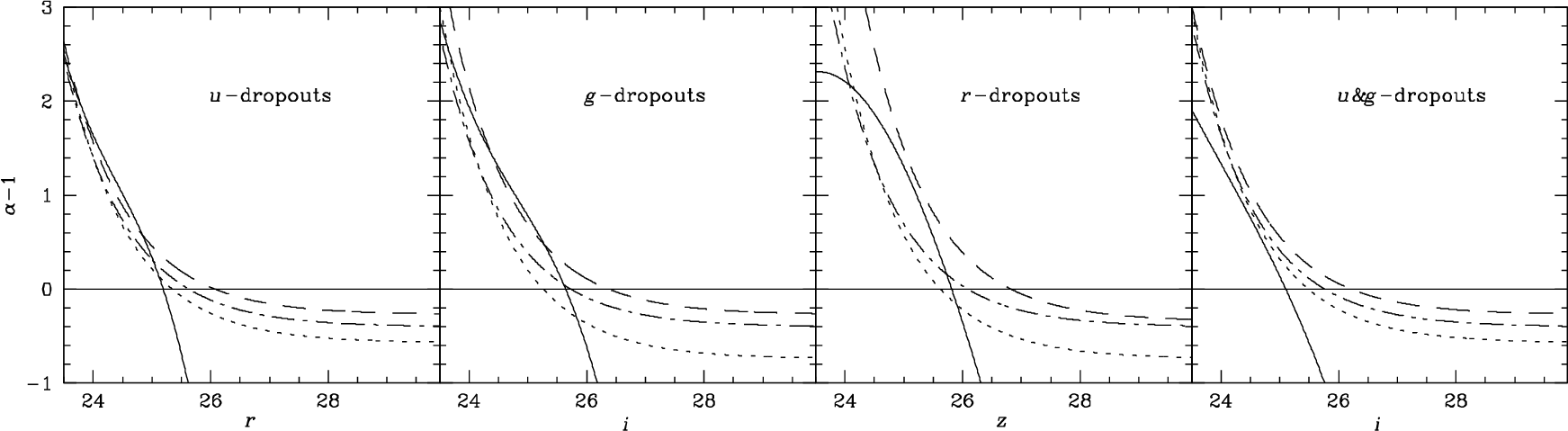}
\caption{Adopted values of $\alpha-1$ as a function of LBG magnitude
  for the four background samples. The \emph{dotted}, \emph{dashed},
  and \emph{dash-dotted} lines correspond to the slopes of the LFs of
  \cite{2006ApJ...642..653S}, \cite{2007ApJ...670..928B}, and
  \cite{1999ApJ...519....1S}, respectively, while the \emph{solid}
  line corresponds to the slope of the polynomial fitted to the number
  counts of Fig.~\ref{fig:numbercounts}.}
\label{fig:alpha}
\end{figure*}

The Schechter function parameters of these external LFs are listed in
Table~\ref{tab:LF}. We fit our number counts with a fourth order
polynomial in magnitude only for comparison to the external LFs and
not for prediction of the magnification signal. In
Fig.~\ref{fig:alpha} we show the adopted values of $\alpha-1$ as a
function of LBG magnitude for the different samples and the different
LF measurements in the literature in comparison to the fitted
polynomial. We use the $z\sim4$ values of \cite{1999ApJ...519....1S}
and \cite{2006ApJ...642..653S} to estimate a $z\sim5$ LF and the
$z\sim4$ values of \cite{2007ApJ...670..928B} to estimate the $z\sim3$
LF assuming no evolution.

\begin{table}
\begin{minipage}[t]{\columnwidth}
\centering

\caption{\label{tab:LF}\cite{1976ApJ...203..297S} function parameters
  for the external LFs} \renewcommand{\footnoterule}{} % to avoid a line before footnotes
\begin{tabular}{lcc}
\hline
\hline
Reference & $M_*$ & $\alpha_{\rm LF}$\footnote{This $\alpha_{\rm LF}$ is the faint-end-slope of the Schechter LF. The $\alpha(m)$ introduced in Eq.~\ref{eq:flux_power_law} approaches this value for faint magnitudes.}\\
\multicolumn{3}{c}{$z\sim3$}\\
\hline
\cite{1999ApJ...519....1S} & $-21.00$ & $-1.60$ \\
\cite{2006ApJ...642..653S} & $-20.90$ & $-1.43$ \\
\cite{2007ApJ...670..928B}\footnote{extrapolated from $z\sim4$} & $-20.98$ & $-1.73$ \\
\hline
\multicolumn{3}{c}{$z\sim4$}\\
\hline
\cite{1999ApJ...519....1S} & $-21.20$ & $-1.60$\footnote{fixed} \\
\cite{2006ApJ...642..653S} & $-21.00$ & $-1.26$ \\
\cite{2007ApJ...670..928B} & $-20.98$ & $-1.73$ \\
\hline
\multicolumn{3}{c}{$z\sim5$}\\
\hline
\cite{1999ApJ...519....1S}$^b$ & $-21.20$ & $-1.60$$^c$ \\
\cite{2006ApJ...642..653S}$^b$ & $-21.00$ & $-1.26$ \\
\cite{2007ApJ...670..928B} & $-20.64$ & $-1.66$ \\
\end{tabular}
\end{minipage}
\end{table}

There is good agreement between the measurement and the literature LFs
at the bright end (the $r$-dropouts suffer from small number
statistics at $z\la 24.5$). However, our measurement of the
faint-end-slope suffers clearly from incompleteness. Thus, whenever
$\alpha-1$ is needed in the remainder of the paper we present results
for these three different external LFs and do not use the number count
slopes measured on our data.

\subsection{Foreground samples}
The foreground samples are selected with the help of photo-$z$'s. In
\cite{2009A&A...498..725H} we showed that we can reach an accuracy of
$\sigma_{z/(1+z)}=0.033$ rejecting only $1.6\%$ of outliers in the
magnitude range $17<i<24$ if we filter for objects with a \emph{BPZ}
ODDS parameter of $\rm ODDS>0.9$. This filter basically rejects
objects with a bimodal redshift-probability function.

Three foreground redshift intervals have been selected: $z=[0.1,1.0]$,
$z=[0.5,1.4]$ and $z=[0.1,0.5]\bigcup [1.0,1.4]$ for cross-correlation
with the $u$-, $g$-, and $r$-dropouts, respectively. For the
analytical predictions, each redshift section is fitted with a
three-parameters $(n_0,z_0,\sigma_z)$ Gaussian function:

\begin{equation}
n(z)=n_0 \exp \left[-\left({z-z_0\over \sigma_z}\right)^2\right]\,.
\end{equation}

\subsection{Clustering measurement}
In the following, we call the catalogue of background LBGs $\rm D_1$
containing $N_{\rm D_1}$ galaxies and the one of the foreground
galaxies $\rm D_2$ with $N_{\rm D_2}$ galaxies. For the measurement of
the cross-correlation function we create one large random catalogue,
called $\rm R$ in the following, with the same areal geometry as the
data catalogues and containing $N_{\rm R}$ objects. We measure the
angular cross-correlation function with a modified version of the
estimator proposed by \cite{1993ApJ...412...64L}:
\begin{equation}
  w(\theta)=\frac{\rm D_1D_2-D_1R-D_2R}{\rm RR}+1\,,
\end{equation}
with $\rm D_1D_2$ being the number of low-$z$-high-$z$ galaxy pairs in
the angular range $[\theta,\theta+\delta\theta]$ normalised by $N_{\rm
  D_1}N_{\rm D_2}$, $\rm D_iR$ being the number of pairs between
catalogue $\rm D_i$ and the random catalogue in that angular range
normalised by $N_{\rm D_i}N_{\rm R}$, and $\rm RR$ being the number of
pairs in the random catalogue in that angular range normalised by
$N_{\rm R}^2$. By choosing $N_{\rm R}\gg N_{\rm D_i}$ (by at least a
factor of ten compared to the largest foreground samples) the shot
noise introduced by the random catalogue can be suppressed. We use
$10^6$ random points for each field that reduce to $\sim7\times10^5$
after masking. The masks used for the masking of the data catalogues
are identical to the ones used for masking the random
catalogues. Halos of bright stars are masked out as well as low-S/N
regions (e.g. the borders of the stack that have lower S/N due to
dithering) and regions affected by diffraction spikes or asteroid
tracks. For a detailed overview of the masking routines we refer the
reader to \cite{2009A&A...493.1197E}. This conservative masking
approach results in a loss of $\sim30\%$ of the area but ensures a
highly uniform dataset with a homogeneous detection and selection
efficiency.

\cite{2003A&A...403..817M} showed that the signal-to-noise of cosmic
magnification measurements can be optimally boosted if an appropriate
weight of $\alpha-1$ is put on each background galaxy. In this way the
sources are weighed according to the expectations from the LF. Bright
LBGs that are expected to be positively cross-correlated to the
low-$z$ lenses because of the steep exponential part of the LF get a
positive weight. Faint LBGs from the shallow part of the LF that are
expected to be anti-correlated get a negative weight. And
intermediately bright LBGs from parts of the LF where
$\alpha-1\approx0$ are down-weighed.  We modify the estimator in the
following way:
\begin{equation}
\label{eq:corr_opt}
  w^{\rm w}(\theta)=\frac{\rm
    D_1^wD_2-D_1^wR-\left<w\right>D_2R+\left<w\right>RR}{\rm RR}\,,
\end{equation}
with $\rm D_1^wD_2$ and $\rm D_1^wR$ being weighted pair counts,
i.e. reflecting the average of the weights of the LBGs in the selected
pairs rather than the pure normalised number of pairs, and $\rm
\left<w\right>$ being the average weight of the LBGs in the whole $\rm
D_1$ catalogue.

We estimate the cross-correlation function separately for each of the
four independent fields and calculate the mean
$\bar w(\theta)$. Furthermore, we draw ten jack-knife samples from the
catalogue of each field and estimate the correlation function for all
40 of these. In order to take the correlation of the errors of
data points at different angular scales properly into account, the
covariance matrix is then estimated in the following way from these
jack-knife samples:
\begin{multline}
  C(\theta_1,\theta_2)=\\\left(\frac{N}{N-1}\right)^2\times\sum_{i}\left[w_i(\theta_1)-\bar w(\theta_1))\times(w_i(\theta_2)-\bar w(\theta_2)\right]\,.
\end{multline}

\section{Results}
\label{sec:results}

\subsection{Cross-correlations in different magnitude bins}

First we cross-correlate LBGs in different magnitude bins to
appropriate (i.e. non-overlapping) low-$z$ samples to see if the
signal agrees with the predictions.

For the $u$-dropouts we choose the low-$z$ range $0.1<z_{\rm
  phot}<1.0$, for the $g$-dropouts we choose $0.5<z_{\rm phot}<1.4$,
and for the $r$-dropouts we choose all galaxies with either
$0.1<z_{\rm phot}<0.5$ or $1.0<z_{\rm phot}<1.4$ (essentially a double
peaked distribution). These choices are motivated by the simulated LBG
redshift distributions shown in Fig.~\ref{fig:z_dist}. We exclude the
low-redshift ranges that potentially contaminate the LBG
samples. Redshift beyond $z=1.4$ are not considered for the lenses
because between $z=1.4$ and $z=2.5$ we cannot expect our photo-$z$'s
to perform very well due to the lack of infrared filters. Furthermore,
we restrict ourselves to magnitudes of $i<24$ for the foreground
sample since without a deeper spectroscopic survey we cannot safely
predict how the rate of catastrophic photo-$z$ outliers develops for
fainter galaxies. We apply an ODDS cut of $\rm ODDS>0.8$ as a
compromise between accuracy and density of the lens samples.

\begin{figure*}
\centering
\resizebox{0.49\hsize}{!}{\includegraphics{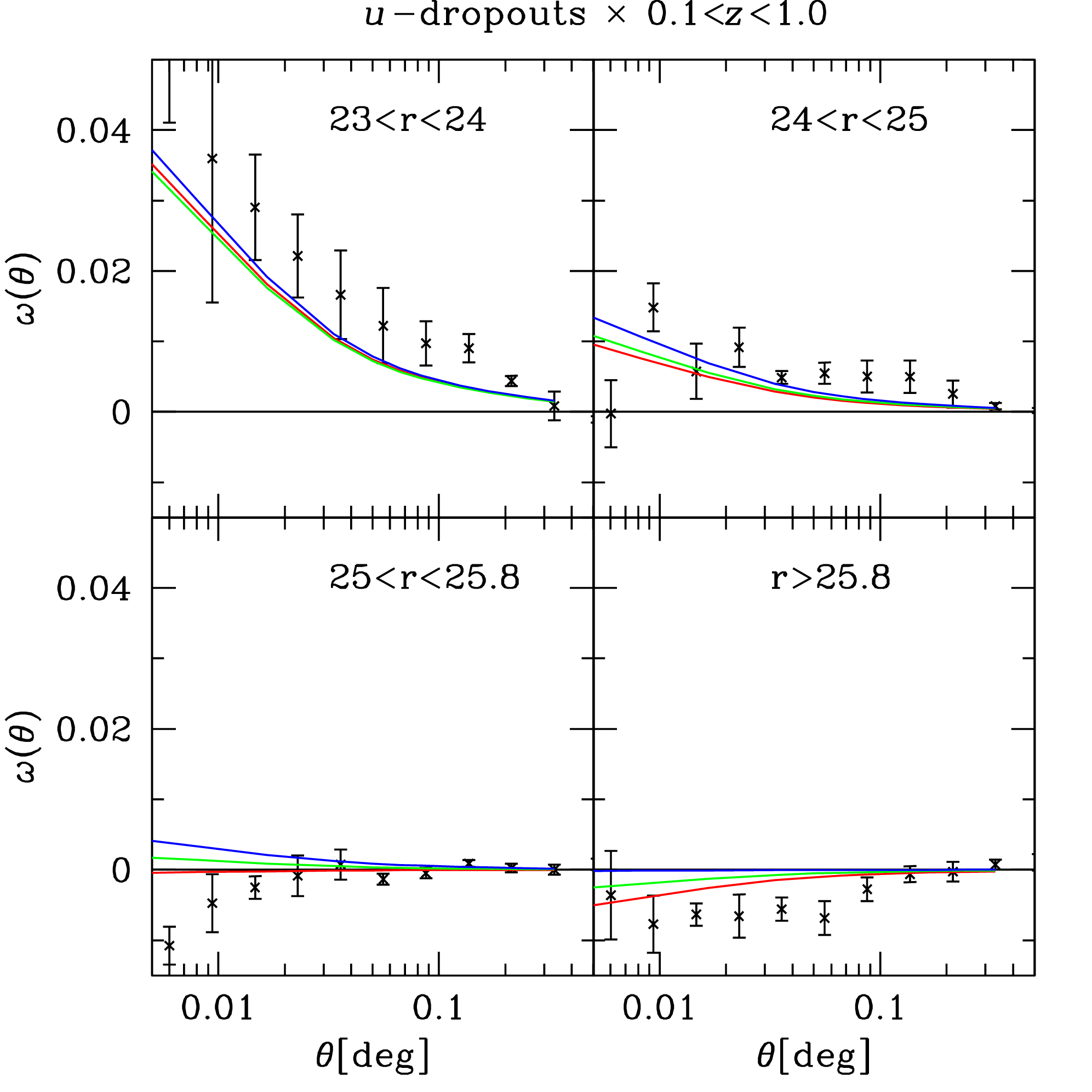}}
\resizebox{0.49\hsize}{!}{\includegraphics{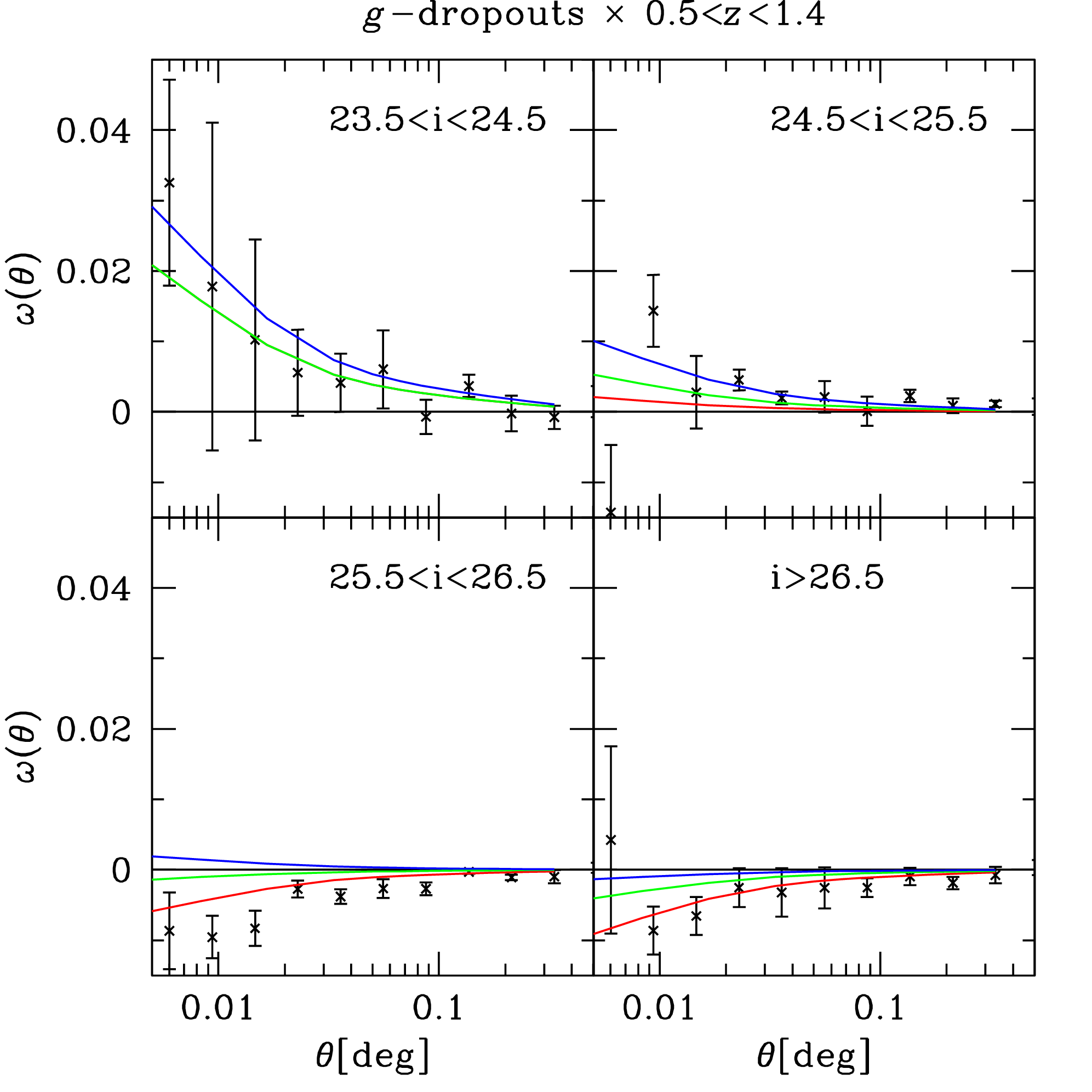}}\\
\resizebox{0.49\hsize}{!}{\includegraphics{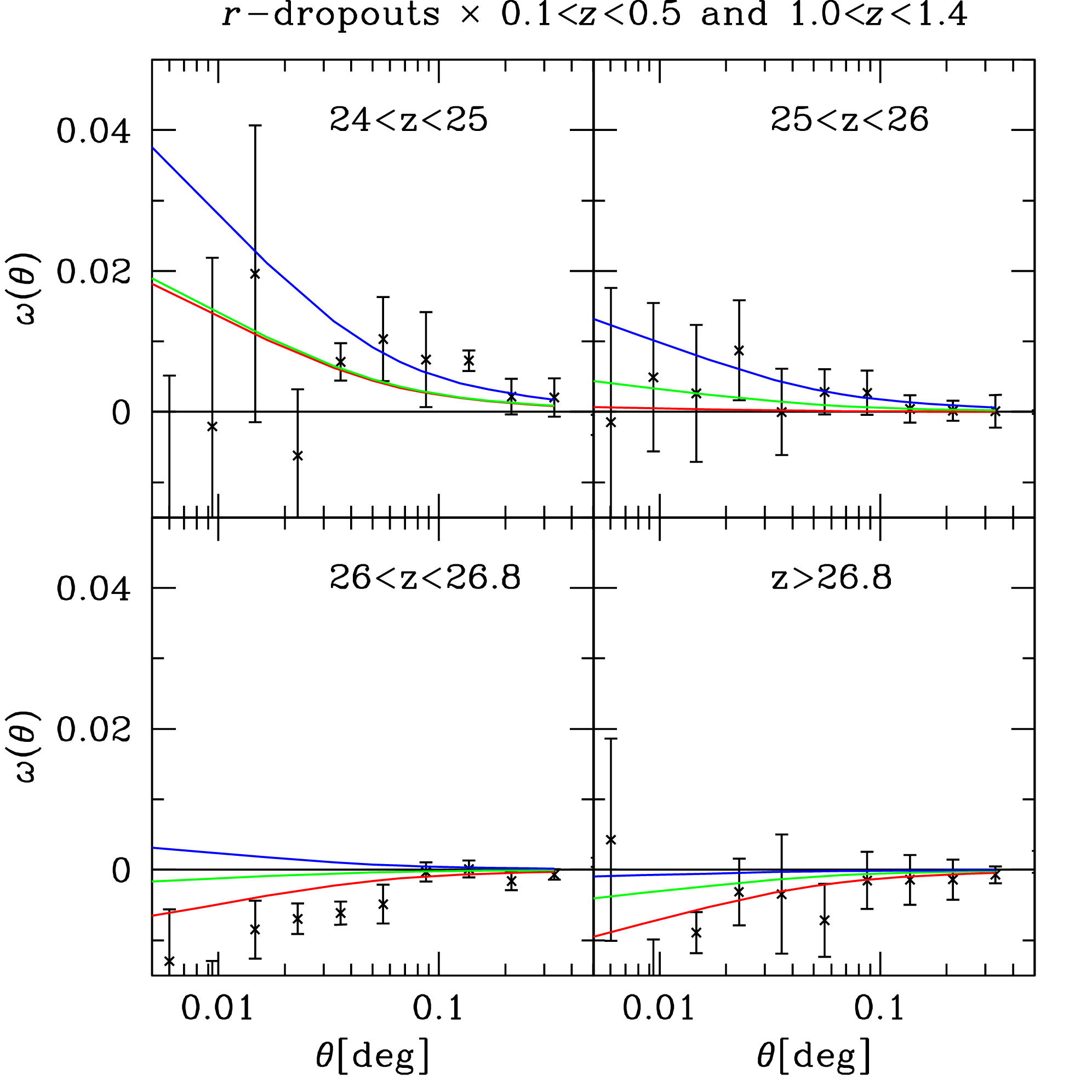}}
\resizebox{0.49\hsize}{!}{\includegraphics{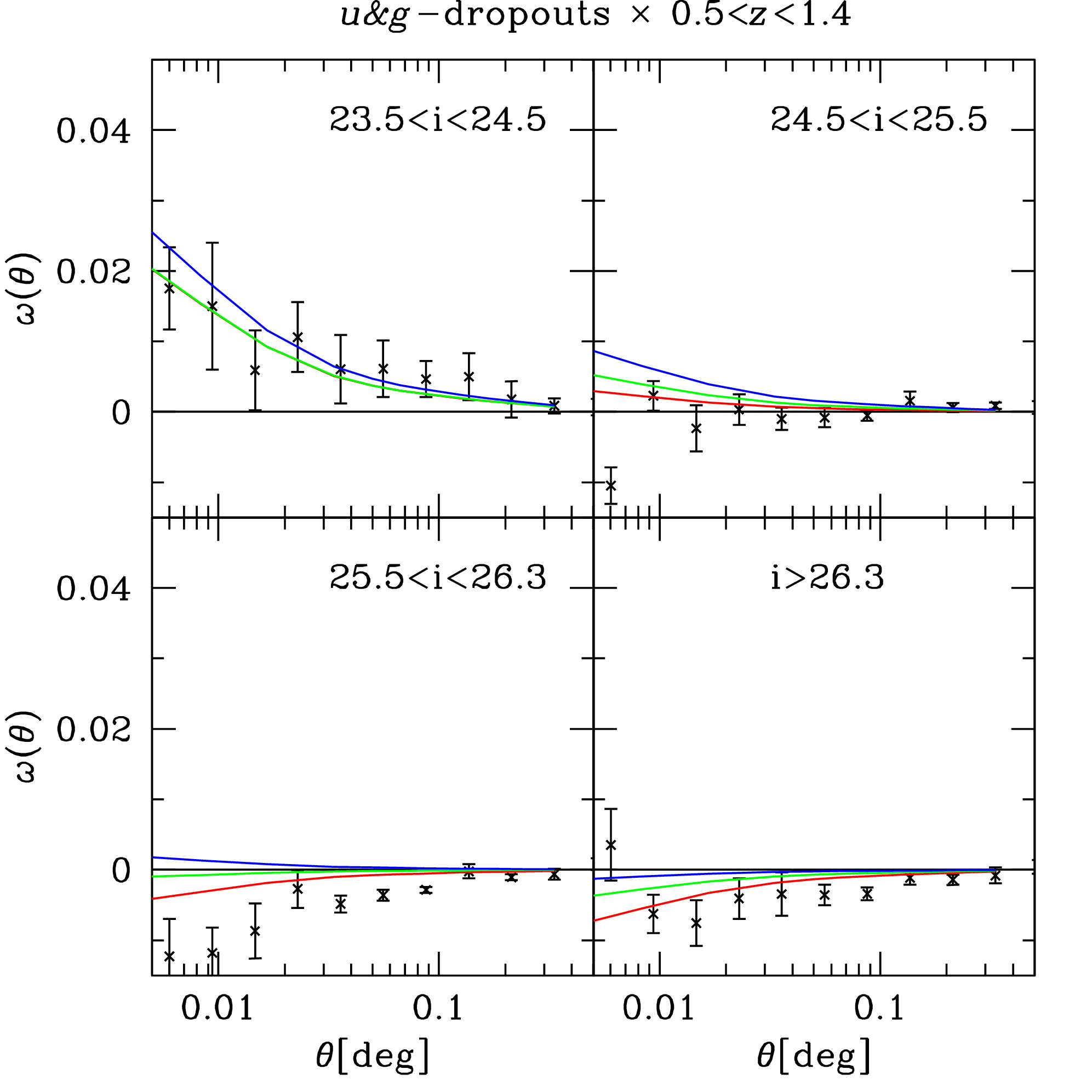}}
\caption{Cross-correlation functions between the dropouts at different
  redshifts and with different magnitudes and the different foreground
  galaxy samples. The red, green, and blue lines correspond to the
  predictions based on the LF slopes of \cite{2006ApJ...642..653S},
  \cite{1999ApJ...519....1S}, and \cite{2007ApJ...670..928B},
  respectively. For some background samples the predictions by
  \cite{2006ApJ...642..653S} and \cite{1999ApJ...519....1S} are
  virtually identical so that the red and green curve lie on top of
  each other.}
\label{fig:cross}
\end{figure*}

In Fig.~\ref{fig:cross} the cross-correlation functions between the
different source samples in different magnitude bins and the lens
samples are shown. Errors are estimated from jack-knife
resampling. The magnitude bins were chosen in such a way that there
are cases with predicted positive ($\left<\alpha-1\right>>0$) and
negative ($\left<\alpha-1\right><0$) amplitudes as well as cases with
an amplitude close to zero. For comparison also the predictions based
on the three different LFs are plotted by using Eq.~\ref{eq:sl_corr}
in combination with the average weight of the LBGs in the particular
magnitude bin, $\left<\alpha-1\right>$.

\subsection{Interpretation of the observed signal}
\label{sec:interpretations}

There is good qualitative agreement between the measured
cross-correlation functions and the predictions from cosmic
magnification. Bright LBGs show a strong positive cross-correlation to
the foreground lenses, intermediately bright LBGs show a signal close
to zero, and the faintest ones are anti-correlated. Especially the
latter observation is a very strong argument for the lensing nature of
the signal, since a physical cross-correlation caused by redshift
overlap could not produce such an anti-correlation.

The magnification predictions are performed from
Eq.~\ref{eq:mu_cross_delta} using the non-linear power spectrum in
\cite{1996MNRAS.280L..19P} with a biasing $b=1$. There is a general
tendency of an underestimation of the signal by the theoretical
predictions. This can have different reasons:

\begin{enumerate}
\item Lensing predictions are performed using the weak lensing
  approximation, i.e $(\kappa ,\gamma)\ll 1$. Next order corrections
  are only of the order of 10\% \citep{2003A&A...403..817M}.
\item The power spectrum normalisation $\sigma_8$ is probably also
  another minor contribution since it is unlikely that the real
  normalisation is very different from the fiducial $\sigma_8=0.8$. 
\item The most likely explanation lies in the biasing parameter $b$ of
  the foreground galaxy population. Further investigation will be
  necessary, in particular the combined analysis with the foreground
  auto-correlation function could remove any direct dependence on $b$
  \citep[][]{Waerbeke2009}.
\end{enumerate}

The predictions based on the LF measurements by
\cite{2006ApJ...642..653S} seem to agree best with our data. The LFs
by \cite{2007ApJ...670..928B} estimated from space-based data with
their very steep faint-end slopes do not yield the negative amplitudes
observed in our cross-correlation functions of the faintest
LBGs. However, we cut the LBG samples in observed magnitudes while
\cite{2007ApJ...670..928B} account for the asymmetric scatter at the
faint end introduced by magnitude errors called Eddington bias
\citep{1913MNRAS..73..359E,2004A&A...424...73T}. The LBGs at the faint
end of our samples have intrinsic magnitudes that are on average
fainter than the observed ones due to this effect. Taking this into
account would lead to more negative values for $\left<\alpha-1\right>$
and theoretical predictions with a larger negative
amplitude. Interestingly, \cite{2006ApJ...642..653S} do not correct
for that asymmetric scatter. We suspect that the better agreement of
our data with the predictions based on their LFs originates from that
fact.

The predictions based on the LF estimates from
\cite{1999ApJ...519....1S} lie in between the predictions from
\cite{2006ApJ...642..653S} and \cite{2007ApJ...670..928B}.

It has also been reported by \cite{2008ApJ...676..767T} that cosmic
variance can lead to a change in the shape of the LF, especially at
the faint end. Pencil-beam surveys in under dense fields tend to yield
steeper slopes than the cosmic average. That may well be another
explanation for the discrepancy between our results and the
predictions based on the HST measurements.

\subsection{Optimally weighted cross-correlation functions}

\begin{figure*}
\centering
\resizebox{0.4\hsize}{!}{\includegraphics{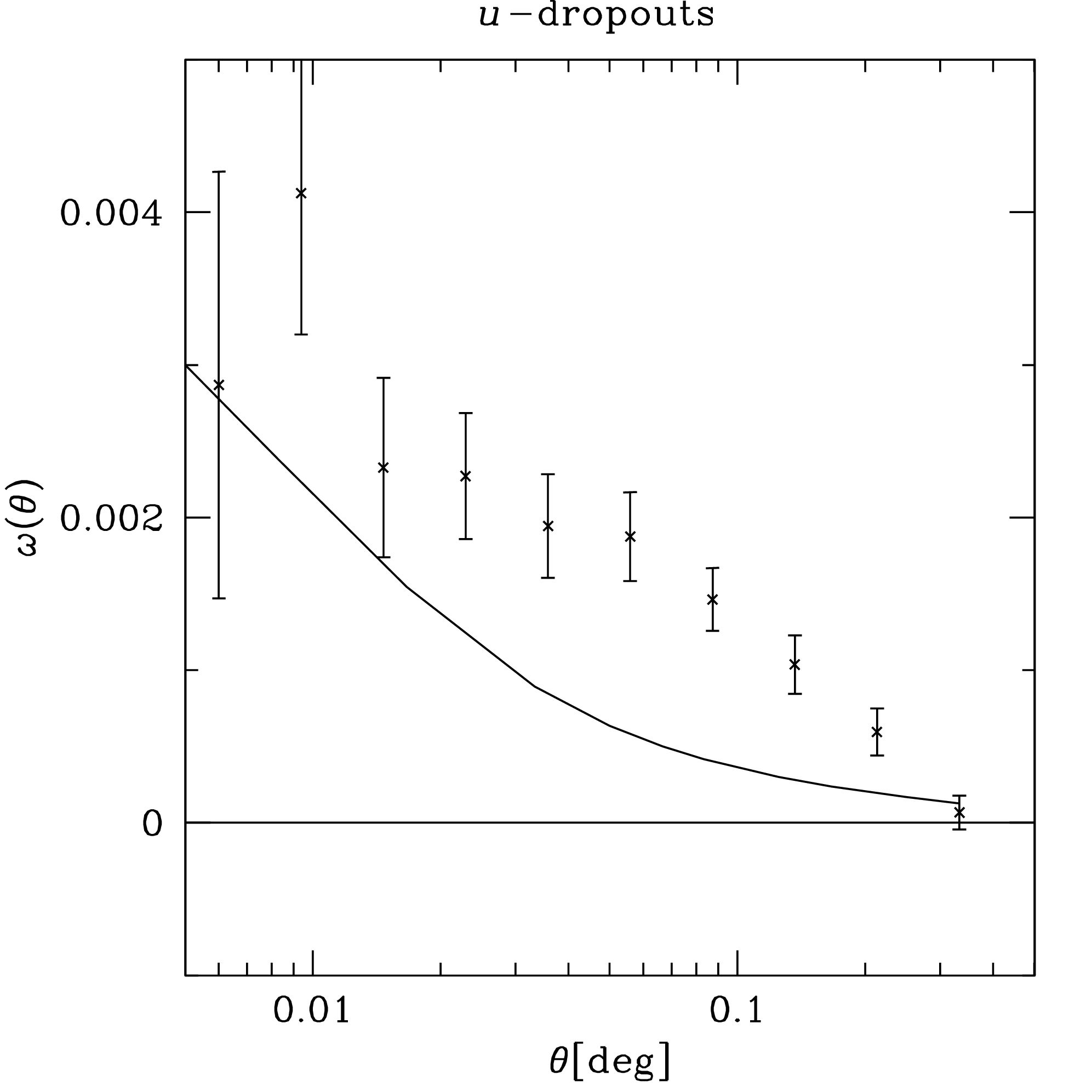}}
\resizebox{0.4\hsize}{!}{\includegraphics{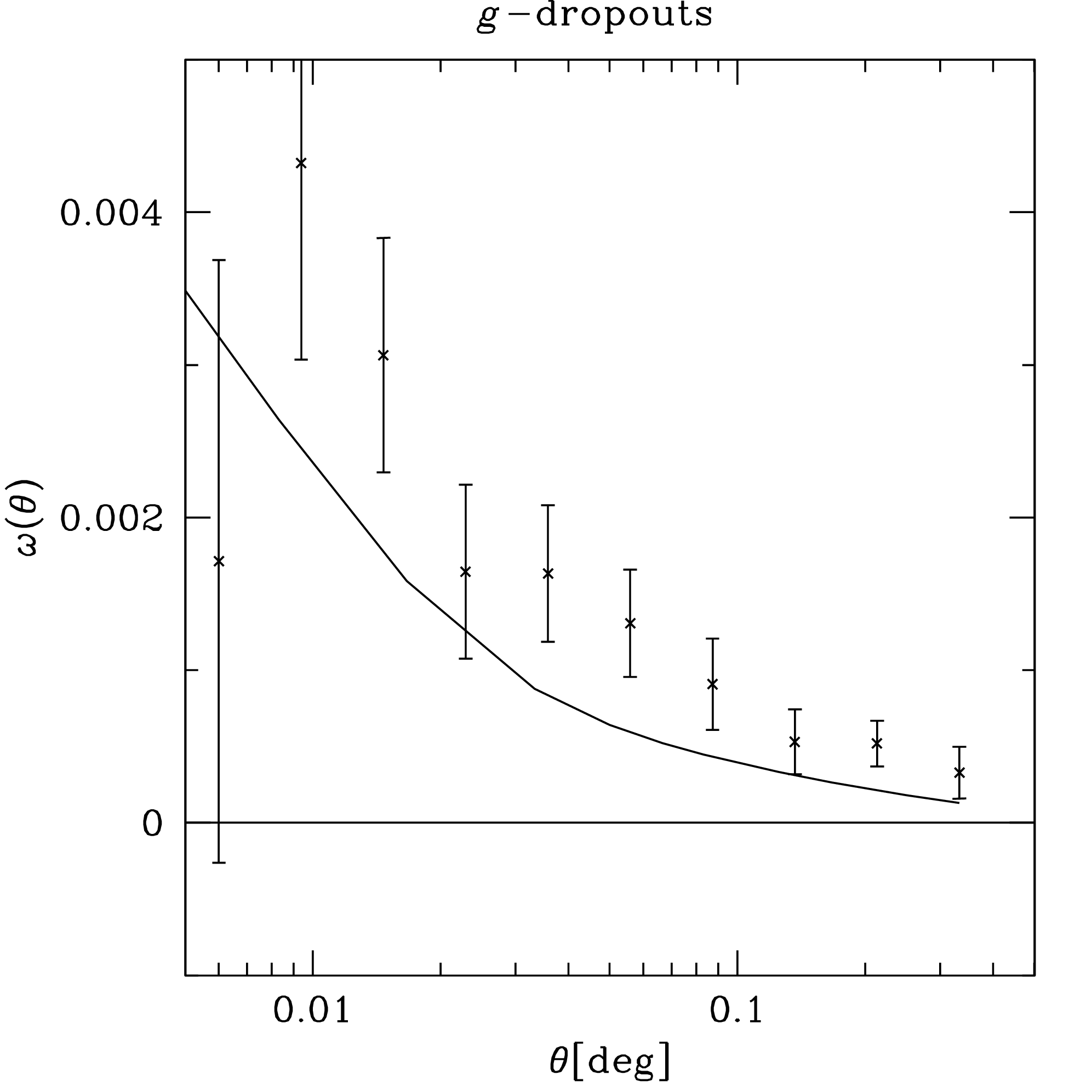}}\\
\resizebox{0.4\hsize}{!}{\includegraphics{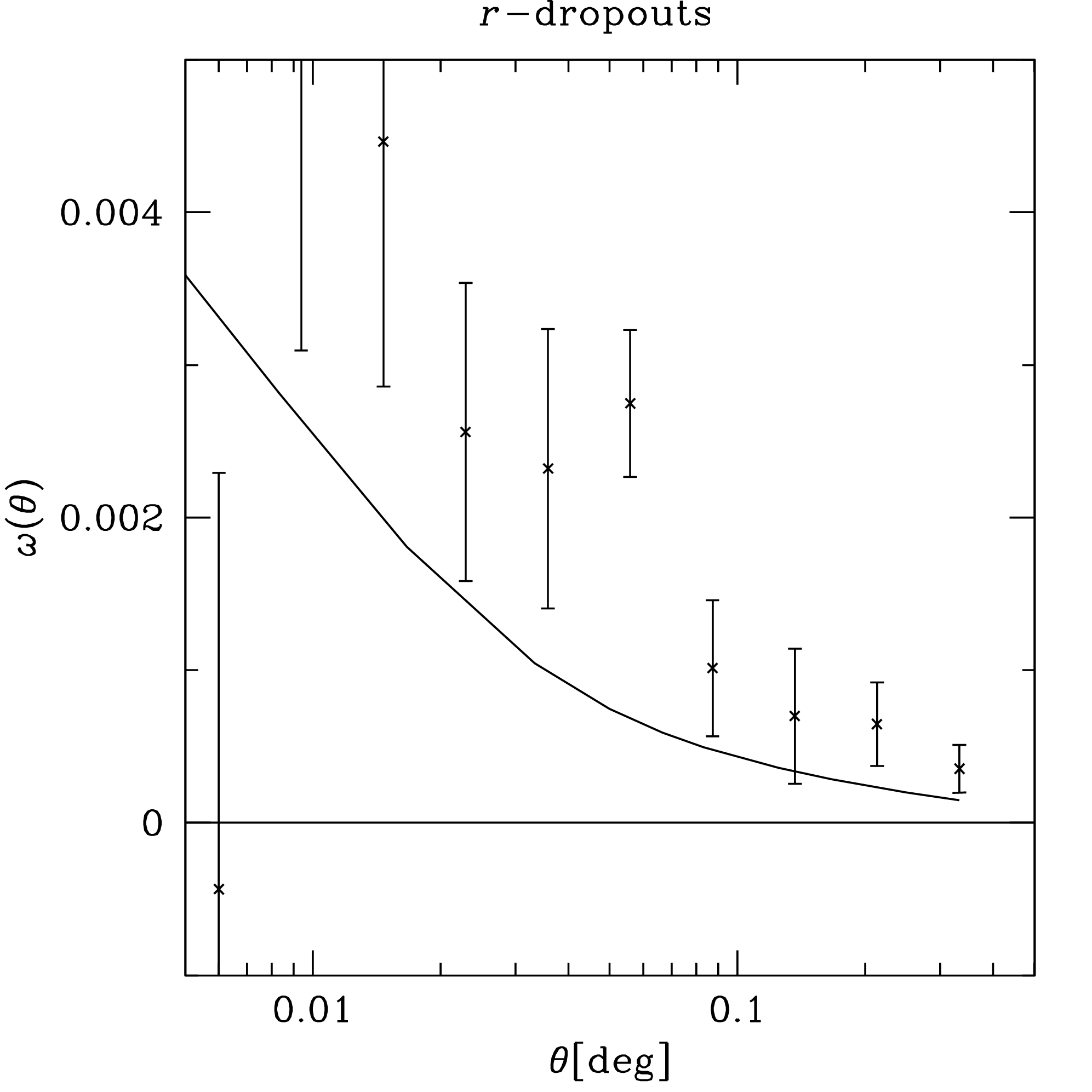}}
\resizebox{0.4\hsize}{!}{\includegraphics{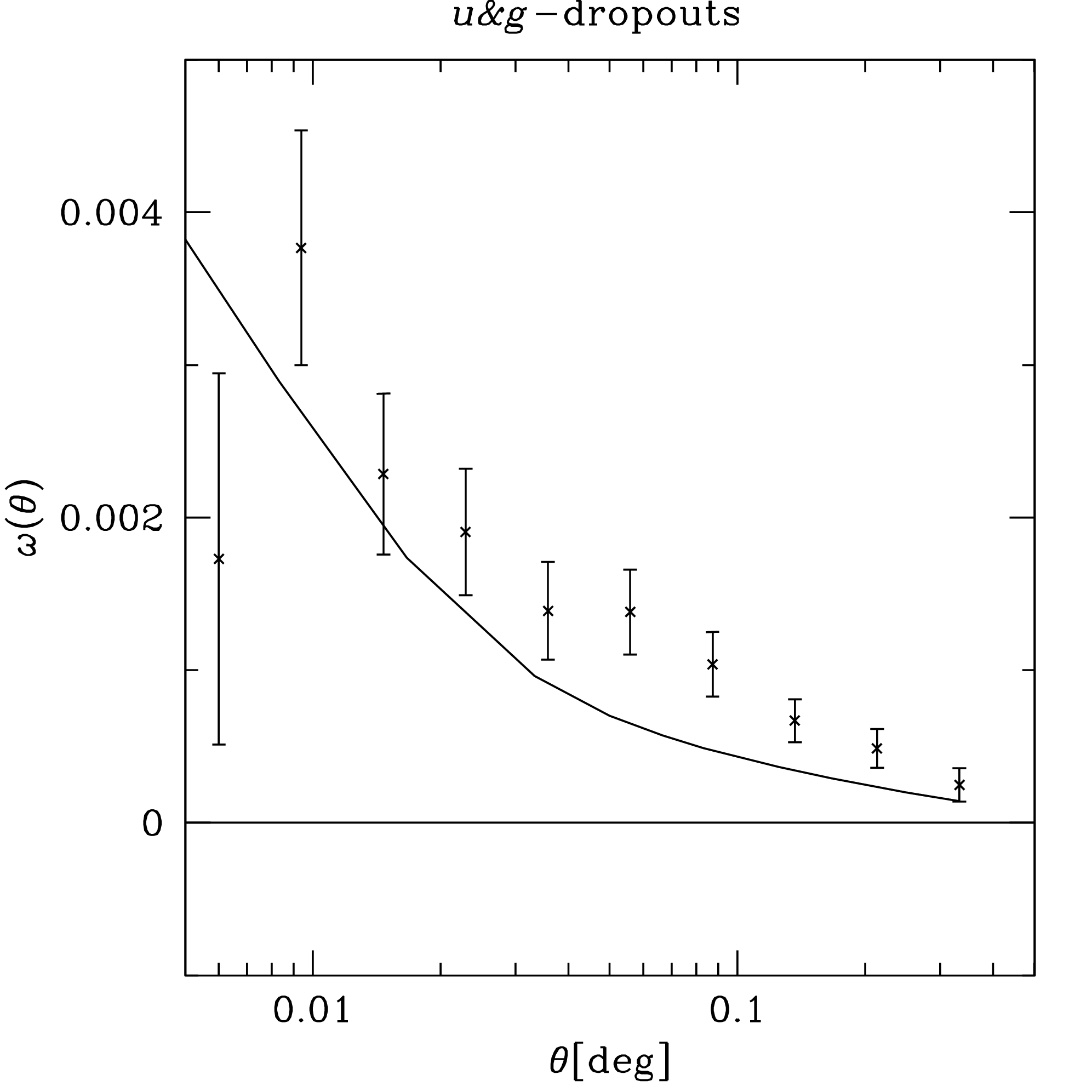}}
\caption{Optimally weighted cross-correlation function between the
  complete dropout samples and the different foreground galaxy
  samples. The solid line correspond to the predictions based on the
  LF slopes of \cite{2006ApJ...642..653S}.}
\label{fig:cross_weight}
\end{figure*}

Next, we estimate optimally weighted cross-correlation functions as
introduced by \cite{2003A&A...403..817M} and also used in
\cite{2005ApJ...633..589S}. We weigh each galaxy with the $\alpha-1$
value corresponding to its magnitude and estimate the correlation
function according to Eq.~\ref{eq:corr_opt}. This is done three times
for the three different sets of LFs. The results for the LFs by
\cite{2006ApJ...642..653S} are displayed in
Fig.~\ref{fig:cross_weight} together with the theoretical
predictions. These are computed with Eq.~\ref{eq:sl_corr} by taking
the average squared weight, $\left<(\alpha-1)^2\right>$, as the
pre-factor.

There is a similar tendency of the theoretical predictions being
slightly lower than the observed signals, most serious for the
$u$-dropouts. The same reasons as discussed in
Sect.~\ref{sec:interpretations} apply here.

\begin{figure}
\centering
\resizebox{\hsize}{!}{\includegraphics[angle=-90]{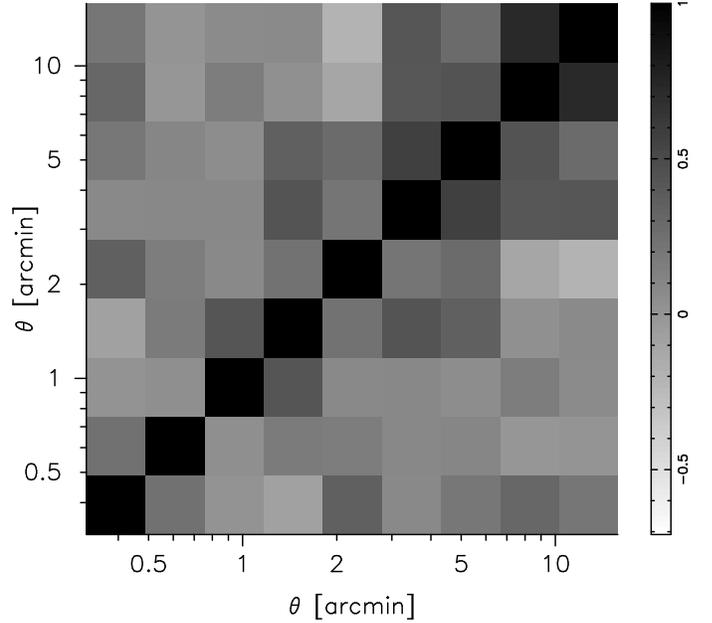}}
\caption{Correlation matrix (normalised covariance matrix) of the
  optimally-weighted cross-correlation function between $u$-dropouts
  and galaxies with $0.1<z_{\rm phot}<1.0$. We display the scales
  $0\farcm3<\theta<15'$ used for the estimation of the total
  significance.}
\label{fig:cov_mat}
\end{figure}

Table~\ref{tab:results} summarises the results for the normal as well
as the weighted correlation functions. For the optimally weighted
cross-correlation functions we report the total significance of the
detection as computed with the help of the covariance matrix. See
Fig.~\ref{fig:cov_mat} for an example of such a matrix.

\begin{table*}
\begin{minipage}[t]{\columnwidth}
\centering
\caption{\label{tab:results}Basic quantities of the samples used in
  the cross-correlation analysis. The subscripts $_{\rm ST}$, $_{\rm
    St}$, and $_{\rm Bo}$ refer to the luminosity function estimates
  of \cite{2006ApJ...642..653S}, \cite{1999ApJ...519....1S}, and
  \cite{2007ApJ...670..928B}, respectively.}
\renewcommand{\footnoterule}{}
\begin{tabular}{lrcccccc}
\hline
\hline
sample & No. of LBGs & $\left<\alpha-1\right>_{\rm ST}$ & $\left<\alpha-1\right>_{\rm St}$ & $\left<\alpha-1\right>_{\rm Bo}$ & $\sigma_{\rm ST}$\footnote{detection significance} & $\sigma_{\rm St}\,^a$ & $\sigma_{\rm Bo}\,^a$\\
\multicolumn{8}{c}{$u$-dropouts}\\
\hline
$23.0 < r < 24.0$ & 714 & 2.03 & 1.97 & 2.14 &  & & \\
$24.0 < r < 25.0$ & 7263 & 0.55 & 0.62 & 0.77 &  & & \\
$25.0 < r < 25.8$ & 15524 & -0.02 & 0.10 & 0.24 &  & & \\
$25.8 < r$ & 10679 & -0.29 & -0.14 & -0.01 &  & & \\
$23.5 < r$ (opt. weights) & 34058 & 0.17\footnote{average squared weight, $\left<(\alpha-1)^2\right>$} & 0.17$^b$ & 0.24$^b$ & 10.21 & 10.47 & 9.93 \\
\hline
\multicolumn{8}{c}{$g$-dropouts}\\
\hline
$23.5 < i < 24.5$ & 990 & 1.28 & 1.28 & 1.78 &  & & \\
$24.5 < i < 25.5$ & 7557 & 0.13 & 0.32 & 0.62 &  & & \\
$25.5 < i < 26.5$ & 18948 & -0.36 & -0.08 & 0.12 &  & & \\
$26.5 < i$ & 8686 & -0.56 & -0.25 & -0.08 &  & & \\
$23.5 < i$ (opt. weights) & 36181 & 0.21$^b$ & 0.10$^b$ & 0.20$^b$ & 6.95 & 7.08 & 5.46 \\
\hline
\multicolumn{8}{c}{$r$-dropouts}\\
\hline
$24.0 < z < 25.0$ & 520 & 1.02 & 1.07 & 2.12 &  & & \\
$25.0 < z < 26.0$ & 3482 & 0.04 & 0.25 & 0.74 &  & & \\
$26.0 < z < 26.8$ & 4768 & -0.37 & -0.09 & 0.18 &  & & \\
$26.8 < z$ & 1685 & -0.53 & -0.23 & -0.05 &  & & \\
$23.5 < z$ (opt. weights) & 10471 & 0.20$^b$ & 0.12$^b$ & 0.51$^b$ & 6.91 & 5.94 & 4.08 \\
\hline
\multicolumn{8}{c}{$u$\&$g$-dropouts}\\
\hline
$23.5 < i < 24.5$ & 5256 & 1.27 & 1.27 & 1.60 &  & & \\
$24.5 < i < 25.5$ & 23212 & 0.18 & 0.33 & 0.54 &  & & \\
$25.5 < i < 26.3$ & 26619 & -0.26 & -0.06 & 0.11 &  & & \\
$26.3 < i$ & 14358 & -0.46 & -0.23 & -0.08 &  & & \\
$23.5 < i$ (opt. weights) & 69445 & 0.24$^b$ & 0.20$^b$ & 0.33$^b$ & 8.98 & 7.24 & 4.96 \\
\end{tabular}
\end{minipage}
\end{table*}

\subsection{Tests for systematics}

In order to test for possible systematics we select stars from our
catalogues via a cut in magnitude and half-light-radius and
cross-correlate these to our LBG samples. The amplitudes of the normal
cross-correlation functions are mostly consistent with zero in all
magnitude bins and for all LBG redshifts. The optimally-weighted
cross-correlation functions are all consistent with zero as well.

Furthermore, we checked the influence of the choice of the foreground
sample. We included galaxies with photo-$z$ estimates in regions where
we would expect some contamination of the LBG samples. For example,
including galaxies with $z_{\rm phot}<0.5$ into the foreground sample
that is cross-correlated to the $g$-dropouts leads to a boost in the
amplitudes. In particular, the anti-correlations, which were observed
before when excluding this low-$z$ range, vanish. The signal turns
positive for the faintest $g$-dropouts. This is in clear contradiction
to the predicted lensing signal which should be negative because of
the shallow slope of the LF at the faint end. Similarly, the negative
signal for the faintest $r$-dropouts turns positive if galaxies with
$0.5<z_{\rm phot}<1.0$ are included in the foreground sample. These
excess signals can be explained by redshift overlap leading to
physical cross-correlations between the small number of contaminants
of the LBG samples and the foreground galaxies.

But then, the fact that we do see negative cross-correlations of the
expected amplitude and angular dependence when these problematic
foreground redshifts are excluded from the low-$z$ sample is a strong
argument for the robustness of the analysis. While a small fraction of
low-$z$ contaminants is probably still present in our background LBG
samples these do not change the amplitude or the shape of the signal
but only add noise since they do not carry a lensing signal.

\section{Conclusions}
\label{sec:conclusions}
For the first time we detect cosmic magnification in samples of normal
galaxies. With the help of the Lyman-break technique we select
background samples of high surface density and large lensing
efficiency (due to their high redshifts) from data of the CFHTLS-Deep
survey. We cross-correlate these LBGs to low-$z$ foreground galaxies
which we select by accurate photo-$z$'s. The expected signals are
estimated by taking external LBG-LF estimates from the
literature. There is good agreement between the observed signals and
the theoretically predicted ones in amplitude as well as in angular
dependence. Some deviations can be explained by Eddington bias, the
linearisation of the magnification, and uncertainties in the
cosmological parameters. The LBG samples used here represent the
highest redshift population that has been used in weak gravitational
lensing so far.

Having proven that cosmic magnification with normal galaxies works in
practice we plan to apply this technique to large imaging surveys in
the future. In contrast to cosmic magnification measurements with
QSOs, using galaxies as sources has the advantage of much higher
source densities. Also compared to cosmic shear and
galaxy-galaxy-lensing this technique might prove competitive since the
number of source galaxies with accurate magnitude measurements and
photo-$z$'s (the only requirements for magnification measurements)
considerably exceeds the number of sources with accurate shape
measurements. Although magnification measurements are less powerful
than shear measurements for a given sample of galaxies, the larger
densities that can be reached with future ground-based surveys will
make precision measurements of cosmic magnification very attractive. A
precise calibration of the LFs of the source galaxies and a robust
correction for observational bias like the Eddington bias is, however,
mandatory to reach that goal.

The cosmological constraints derived from cosmic magnification
measurements will be complementary to other probes such as cosmic
shear because their dependence on redshift is slightly different. This
can potentially lead to breaking degeneracies in cosmological
parameters. See also the theoretical companion paper by
\cite{Waerbeke2009} about this topic. Furthermore, cosmic
magnification depends on completely different systematics. Measuring
the same cosmological quantity with e.g. cosmic shear and cosmic
magnification simultaneously will be an extremely important
consistency check. Thus, cosmic magnification and cosmic shear can
become a powerful combination in unravelling the mysteries of dark
matter and dark energy.

\begin{acknowledgements}
We would like to thank P. Simon for support with his correlation
function code, R. Scranton for helpful comments about the data
analysis, and J. Hartlap \& T. Schrabback for help with some of the
plots. We are grateful to the CFHTLS survey team for conducting the
observations and the TERAPIX team for developing software used in this
study. We acknowledge use of the Canadian Astronomy Data Centre
operated by the Dominion Astrophysical Observatory for the National
Research Council of Canada's Herzberg Institute of Astrophysics. HH
would like to thank UBC, Vancouver, for great hospitality and the
European DUEL RTN, project MRTN-CT-2006-036133, for support. LVW was
supported by the Canadian Foundation for Innovation, NSERC and
CIfAR. This work was supported by the DFG priority program SPP-1177
"Witnesses of Cosmic History: Formation and evolution of black holes,
galaxies and their environment" (project ID ER327/2-2), the German
Ministry for Science and Education (BMBF) through DESY under the
project 05AV5PDA/3 and the TR33 "The Dark Universe".
\end{acknowledgements}

\bibliographystyle{aa}

\bibliography{cosmic_magnification}

\end{document}